\documentclass[fleqn,10pt]{wlscirep}

\usepackage{hyperref}
\usepackage{breakurl}
\hypersetup{pdftitle=DebtRank: A microscopic foundation for shock propagation, pdfauthor={Marco Bardoscia, Stefano Battiston, Fabio Caccioli, Guido Caldarelli}}

\title{DebtRank: A microscopic foundation for shock propagation}

\author[1,*]{Marco Bardoscia}
\author[2,1]{Stefano Battiston}
\author[3]{Fabio Caccioli}
\author[4,1,5]{Guido Caldarelli}

\affil[1]{London Institute for Mathematical Sciences, London, W1K 2XF, United Kingdom}
\affil[2]{University of Z\"{u}rich, Department of Banking and Finance, Z\"{u}rich, 8032, Switzerland}
\affil[3]{University College London, Department of Computer Science, London, WC1E 6BT, United Kingdom}
\affil[4]{Institute for Advanced Studies, Lucca, 55100, Italy}
\affil[5]{CNR-ISC: Institute for Complex Systems, Rome, 00185, Italy}
\affil[*]{mb@lims.ac.uk}


\begin{abstract}
The DebtRank algorithm has been increasingly investigated as a method to estimate the impact of shocks in financial networks, as it overcomes the limitations of the traditional default-cascade approaches. 
Here we formulate a dynamical ``microscopic'' theory of instability for financial networks by iterating balance sheet identities of individual banks and by assuming a simple rule for the transfer of shocks from borrowers to lenders. 
By doing so, we generalise the DebtRank formulation, both providing an interpretation of the effective dynamics in terms of basic accounting principles and preventing the underestimation of losses on certain network topologies.
Depending on the structure of the interbank leverage matrix the dynamics is either stable, in which case the asymptotic state can be computed analytically, or unstable, meaning that at least one bank will default. We apply this framework to a dataset of the top listed European banks in the period 2008 - 2013. We find that network effects can generate an amplification of exogenous shocks of a factor ranging between three (in normal periods) and six (during the crisis) when we stress the system with a 0.5\% shock on external (i.e.\ non-interbank) assets for all banks. 
\end{abstract}

\begin{document}

\flushbottom
\maketitle
\thispagestyle{empty}

\section*{Introduction}
The recent economic downturn has made clear that some substantial features of the present financial markets have not been properly considered. Regulators\cite{haldane2009banks,haldane2009rethinking,trichet2010opening} and 
academics\cite{cont2010network} pointed out the role played by complexity\cite{Haldane2011,battiston2013complex,caccioli2009eroding} in the little understanding of the crisis, and in particular the lack of a quantitative assessment for the level of interconnectedness.
It has been increasingly recognised that the main and simplest way to quantitatively account for the degree of interconnectedness and complexity of financial markets is given by the theoretical framework of complex networks \cite{caldarelli2007scale,boss2004network,demasi2006fitness,soramaki2007topology}. 
By representing financial institutions as vertices of a graph we can identify the systemically important ones with the most central vertices\cite{eisenberg2001systemic,elsinger2006risk,Nier2007}.
Furthermore, the evolution of systemic risk can also be modelled by means of dynamical processes on networks \cite{miranda2013contagion,martinez2014empirical,Markose2012,montagna2014contagion,battiston2012debtrank,brummitt2014inside}.  

On the one hand, the use of networks makes the quantification and visualisation of interconnectedness possible; on the other hand, and perhaps even more importantly, network effects are also responsible for a more subtle, typically unnoticed but crucial effect: the amplification of distress. 
Indeed, while diversification archived through a higher level of interconnectedness reduces the individual risk (in the case of independent shocks), it can however increase systemic risk \cite{caccioli2014stability,caccioli2015overlapping,corsi2013micro,bardoscia2012financial,battiston2014systemic}. Nevertheless, there is no single topological structure that is the most robust in all situations because market liquidity also matters \cite{roukny2013default}.
All these issues are presently considered by regulators\cite{boe2013stresstesting} and the notion of interconnectedness has already entered the  debate on ``Global Systemically Important Banks'' (G-SIBs) \cite{basel2011global}.

When the banking system is represented as a network, usually propagation of shocks takes place only with removal of vertices in the system, i.e.\ only after default events. This is an important mechanism for contagion between counterparties\cite{Nier2007,Gai2010,Upper2011,caccioli2012heterogeneity}, although in practice this channel becomes active only if balance sheets are already quite deteriorated \cite{martinez2014empirical} or in combination with other contagion channels, such as those due to fire sales and overlapping portfolios \cite{halaj2013assessing,hurd2011framework,caccioli2014stability,caccioli2015overlapping}. The DebtRank algorithm \cite{battiston2012debtrank} was introduced precisely to overcome this limitation, and to account for the incremental build-up of distress in the system, even before the occurrence of defaults.

At the ``microsocopic'' level, every financial institution satisfies a balance sheet identity that links the values of its assets and liabilities to a capital buffer, which is meant to absorb losses. Balance sheets of different banks are interconnected and therefore the mutual interaction between them is expected to play a major role in the emergence of collective properties, as it is usually the case for many diverse complex systems. For example, our result for the stability of the system, i.e.\ that it depends only on structural properties and not on the initial state, is a clear example of a general property that finds applications in different domains.

The original DebtRank\cite{battiston2012} helped to shift the attention towards interconnectedness as a crucial driver of systemic risk\cite{Thurner2013}. In this paper we show that a similar dynamics can be derived from basic accounting principles and from a simple mechanism for the propagation of shocks from borrower banks to lender banks. 
A limitation of the original DebtRank is that banks pass on distress to their creditors only once, leading in some cases to a significant underestimation of the level of distress in the system. The dynamics proposed here overcomes this limitation by allowing further propagations of shocks.
Perhaps the most important point is that we are able to characterise the qualitative behaviour of the system by establishing a crucial link between the stability of our dynamics and the largest eigenvalue of the interbank leverage matrix. 
One of the hallmarks of DebtRank is that it allowed regulators to monitor at the same time \emph{impact} and \emph{vulnerability} of financial institutions by quantifying in terms of monetary value the impact of the received shocks. Hence, we test our algorithm on a dataset of $183$ European banks listed on the stock market. Our analysis shows that systemic risk has consistently decreased between 2008 and 2013, and that banks having the largest impact on the system are also the most vulnerable ones.

\section*{Results}

\subsection*{Model description}
We represent the interbank system as a directed network whose nodes are banks. A link of weight $A_{ij}$ from node $i$ to node $j$ corresponds to an interbank loan from the lender bank $i$ to the borrower bank $j$ of amount $A_{ij}$ USD. As such, every node is characterised by an internal structure given by its balance sheet (see Methods). On the asset (liability) side we distinguish between interbank and external assets (liabilities). The interbank assets of bank $i$ correspond to the total amount of outstanding loans to other banks within the system, i.e.\ $\sum_j A_{ij}$, while non-interbank assets are called external assets and denoted by $A^E_i$. For every interbank asset $A_{ij}$ in the balance sheet of bank $i$ there is a corresponding interbank liability $L_{ij}=A_{ij}$ in the balance sheet of bank $j$. As a consequence, links can be interpreted as connections between specific elements of balance sheets, i.e.\ of nodes internal structure. Each bank $i$ also has external liabilities $L^E_{i}$, which correspond to obligations to entities outside the system. The equity $E_i$ of bank $i$ is defined through the balance sheet identity as the difference between its total assets and liabilities. We say that bank $i$ has defaulted if $E_i\leq 0$, i.e. if its total liabilities exceeds its total assets. This is in fact only a proxy for a real default event, which is however a common assumption in the literature on financial contagion (see for instance \cite{Nier2007,Gai2010,Upper2011,caccioli2012heterogeneity}). 

We now want to write an equation for the evolution of the equity of all banks which remains consistent with the balance sheet identity over time.
We first define the set of active banks at time $t$ as the set of banks that have not defaulted up to time $t$:
\begin{equation} \label{eq:active_nodes}
\mathcal{A}(t) = \{ j : E_j(t) > 0 \} \, .
\end{equation}
In the following, we will consider a mark-to-market valuation for interbank assets, while liabilities will keep their face value. The idea behind this assumption is that the effect of a bank $j$ being under distress is almost immediately incorporated into the value of the interbank assets $A_{ij}$ held by a creditor bank $i$, while the obligations of bank $j$ to bank $i$ do not change. When bank $j$ defaults, it defaults on all its interbank liabilities, meaning that its creditors will not recover the money that was lent to $j$ and $A_{ij}$ will be zero. As a consequence, the balance sheet identity for bank $i$ at time $t$ reads:
\begin{equation} \label{eq:bs_identity}
E_i(t) = A_i^E(t) - L_i^E(t) + \sum_{j \in \mathcal{A}(t-1)} A_{ij}(t) - \sum_{j=1}^N L_{ij}(t) \, . 
\end{equation}
The reason why the sum involving interbank assets runs over all banks active at time $t-1$ is that the information about the default of other banks is received by bank $i$ with a delay, and accounted for only at the next time step.

We next assume a simple mechanism for shock propagation from borrowers to lenders. The idea is that relative changes in the equity of borrowers are reflected in equal relative changes of interbank assets of lenders at the next time-step:
\begin{equation} \label{eq:shocks}
A_{ij}(t+1) = 
\begin{cases} 
A_{ij}(t) \frac{E_j(t)}{E_j(t-1)} &\mbox{if } j \in \mathcal{A}(t-1) \\
A_{ij}(t) = 0 &\mbox{if } j \notin \mathcal{A}(t-1) \, ,
\end{cases}
\end{equation}
where the case $j \notin \mathcal{A}(t-1)$ ensures that, once bank $j$ defaults, the corresponding interbank assets $A_{ij}$ of its creditors will remain zero for the rest of the evolution. Suppose, for example, that bank $j$ defaults at time $s$, i.e.\ $E_j(s-1) > 0$, but $E_j(s) = 0$; as a consequence, $A_{ij}(s+1) = 0$, for all $i$. At time $s+2$, since $j \notin \mathcal{A}(s)$, the second case will apply, and $A_{ij}(s+2) = 0$. For $t > s+2$, obviously, $A_{ij}(t)$ will remain equal to zero. 

By iterating the balance sheet identity \eqref{eq:bs_identity} and the shock propagation mechanism \eqref{eq:shocks}, the contagion dynamics can be conveniently cast (see Methods for a detailed derivation) in terms of the relative cumulative loss of equity for bank $i$: $h_i(t) = \left( E_i(0) - E_i(t) \right) / E_i(0)$:
\begin{subequations} \label{eq:dyn}
\begin{equation} \label{eq:evo_h}
h_i(t+1) = \min \left[ 1, \, h_i(t) + \sum_{j=1}^N \Lambda_{ij}(t) \left[ h_j(t) - h_j(t-1) \right] \right] \, ,
\end{equation}
\begin{equation} \label{eq:leverage}
\Lambda_{ij}(t) = 
\begin{cases} 
\frac{A_{ij}(0)}{E_i(0)} &\mbox{if } j \in \mathcal{A}(t-1) \\
0 &\mbox{if } j \notin \mathcal{A}(t-1) \, ,
\end{cases}
\end{equation}
\end{subequations}
where we call $\Lambda$ the interbank leverage matrix.

The above dynamics resembles the DebtRank algorithm already introduced in the literature \cite{battiston2012debtrank}. 
An important difference is that in the original DebtRank a bank is allowed to propagate shocks only the first time it receives them. In some cases this might lead to a severe underestimation of the losses. Let us suppose that bank $i$ is hit at time $t$ by a small shock, which will be propagated resulting in additional small shocks at time $t+1$ for its creditors. If the network does not contain any loop bank $i$ will not be hit again by any other shock. However, if the network does contain loops bank $i$ might be hit at later times by a shock which, depending on how much leveraged its borrowers are, might be far larger than the first one, but it will be unable to propagate it. Eq.\ \eqref{eq:dyn} is more general in the sense that as long as a bank receives shocks it will keep propagating them. In fact it can be proved that the two algorithm give the same losses on a certain class of networks (as trees), but, in general, the losses computed via the original DebtRank are a lower bound to those computed with \eqref{eq:dyn}. More precisely, if we shock a single node $s$, the two algorithms will give the same losses for all nodes $r$ such that a unique path from $r$ to $s$ exist. If we shock more nodes, the two algorithms will give the same losses for all nodes $r$ such that unique and non-overlapping paths between $r$ and all the shocked nodes exist. On all the other cases \eqref{eq:dyn} leads to larger losses (see Methods).

A crucial feature of the dynamics \eqref{eq:dyn} is that its stability is determined by the properties of the interbank leverage matrix $\Lambda(t)$. Notably, it is possibile to show (see Methods) that when $|\lambda_{\text{max}}|$, the modulus of the largest eigenvalue of $\Lambda(t)$, is smaller than one, the dynamics converges to the fixed point $\Delta h(t) \equiv h(t) - h(t-1) = 0$, meaning that the shock is progressively damped in subsequent rounds. In contrast, when $|\lambda_{\text{max}}| > 1$ the initial shock will be amplified and at least one bank will default. Remarkably, this happens independently on the properties of the initial shock. After the default, according to \eqref{eq:leverage}, $\Lambda(t)$ will be modified and the same argument will apply to the new interbank leverage matrix. The dynamics will eventually converge when the modulus of the largest eigenvalue of $\Lambda(t)$ becomes smaller than one. 
This explains why, even if the system is initially in the unstable phase, the dynamics does not necessarily converge to the state in which all the banks default. When a bank defaults it is effectively removed from the system when the interbank leverage matrix is updated. The new, \emph{reduced} system could now be in stable phase, and thus converge to the stable fixed point. 
The important point here is that, although the exact values of final losses will depend on the initial shock, the ability of the system to amplify distress and lead to defaults is an exclusive property of the leverage matrix. 
This result confirms the importance of the leverage matrix for the amplification of shocks within the context of systemic stability, as suggested by \cite{Markose2012}, albeit for a different contagion mechanism (the so-called Furfine algorithm \cite{furfine2003quantifying}).

\subsection*{Application to the European banking system}
We now apply the introduced algorithm to the European banking system. We use data from the balance sheets of 183 publicly traded European banks between 2008 and 2013. Available data only contain information about the total amount of interbank borrowing and lending for each bank, which are respectively the sum over rows and columns of the matrix of interbank assets $A_{ij}$. Therefore, we resort to a two-steps reconstruction technique \cite{Battiston2015,Cimini2014a,Cimini2014b} to infer plausible values for all the entries of the matrix. In the first step we build the topology of the network using a so-called fitness model, while in the second step we assign weights to links using the RAS algorithm \cite{upper2004estimating} (see Methods for more details about the reconstruction procedure). Due to the stochasticity of the first step, we sample 100 different networks, which will be used in the following experiments.

\begin{figure}[h]
\centering
\includegraphics[width= \columnwidth]{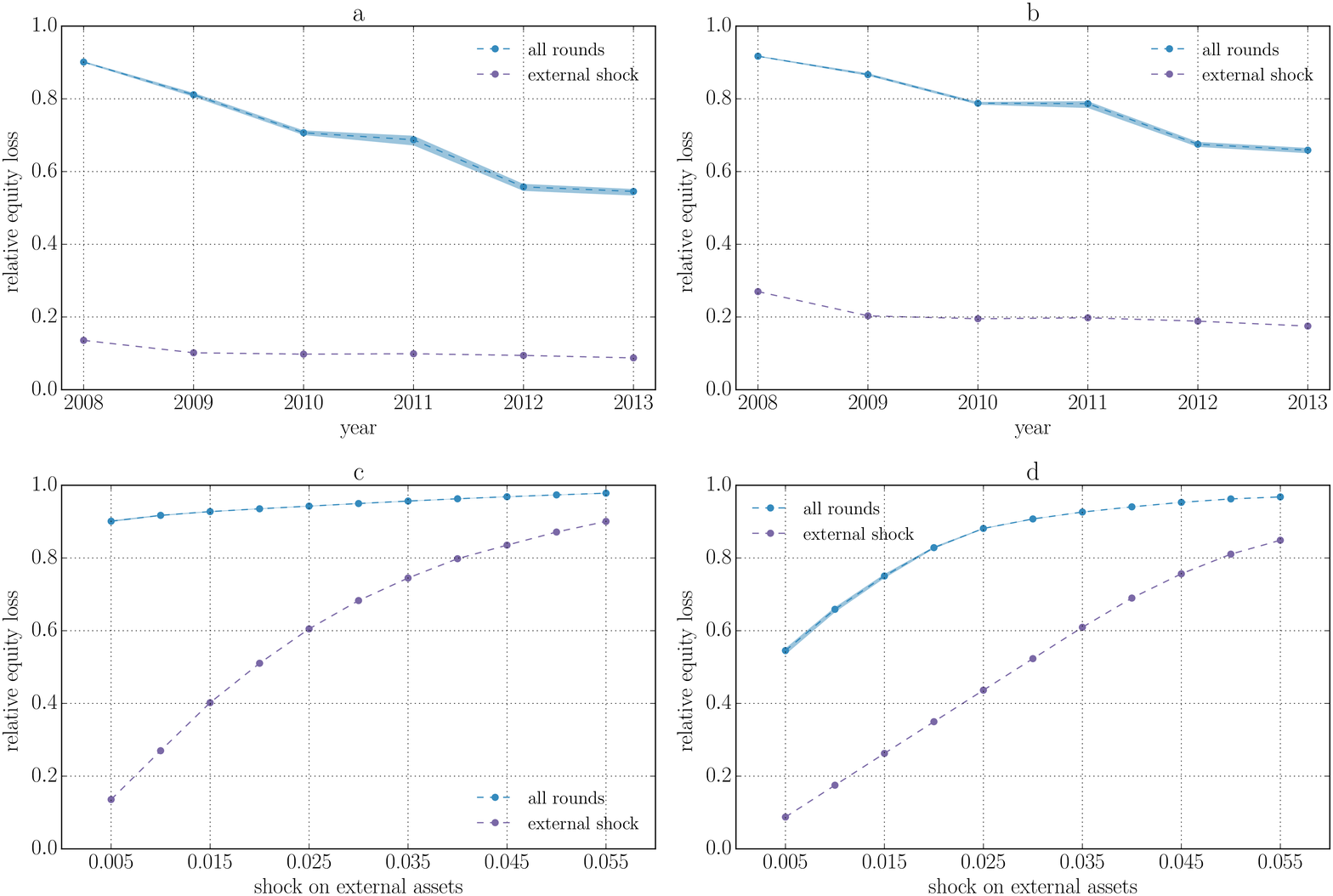}
\caption{\textbf{Relative equity loss for the system of $183$ publicly traded EU banks between 2008 and 2013.} All banks are subject to an initial shock consisting in the devaluation of their external assets by a factor $\alpha$. The violet curves represent the relative equity loss that is directly due to the initial shock, while the blue curves include losses due to the contagion dynamics. Every point is the average over 100 reconstructed networks and the semi-transparent region covers the range between the minimum and maximum across the sample. $\alpha$ is fixed in the top panels and equals to $0.5\%$ (panel a) or $1\%$ (panel b). We see that the amplification effect is reduced from 2008 to 2013. Bottom panels refer to 2008 (panel c) and 2013 (panel d). We see that the relative equity loss saturates for large enough shocks. In 2008 the saturation already occurs for shocks as large as $0.5\%$.}
\label{fig:rel}
\end{figure}

As a first scenario, we consider a shock affecting all banks simultaneously at time $t = 1$ corresponding to a relative devaluation $\alpha$ of their external assets. Following \cite{Battiston2015} we measure the response of each bank to the shock in terms of its contribution $H_i(t)$ to the relative equity loss of the system:
\begin{equation}
H_i(t) \equiv \frac{E_i(0) - E_i(t)}{\sum_i E_i(0)} = h_i(t) \frac{E_i(0)}{\sum_i E_i(0)} \, .
\end{equation}
The direct effects of the shock in terms of relative equity loss are $H_i(1)$, while the effects of contagion are computed using the algorithm introduced here, which is run until convergence (see Methods for more details). In the top panels of Fig.\ \ref{fig:rel} we compare the total relative equity loss $H(t) = \sum_i H_i(t)$ directly due to the initial shock (i.e.\ at time $t = 1$) with the one that includes losses generated by the contagion dynamics (i.e.\ at the convergence of the algorithm), for all the years, and for $\alpha = 0.5\%$ and $1\%$. The overall behaviour resembles the one reported in \cite{Battiston2015} obtained using the original DebtRank. However, as already discussed, the relative equity losses observed here are larger by a factor ranging from $1.3$ in 2008 to $1.7$ in 2013.

We further test this scenario in the bottom panels of Fig.\ \ref{fig:rel} by focusing on 2008 and 2013 and letting $\alpha$ vary between 0.5\% and 5.5\%. The relative equity loss experienced by the system increases as we increase $\alpha$, until it reaches a saturation point. For large enough values of $\alpha$ most of the equity of banks is already wiped out by the initial shock, implying that the amplification due to the contagion dynamics decreases with $\alpha$. Interestingly, we observe that in 2008 the amplification pushes relative losses of equity to saturation levels already for values of $\alpha$ as small as 0.5\%, while in 2013 shock five times larger are needed to reach similar relative losses.

\begin{figure}[h]
\centering
\includegraphics[width= \columnwidth]{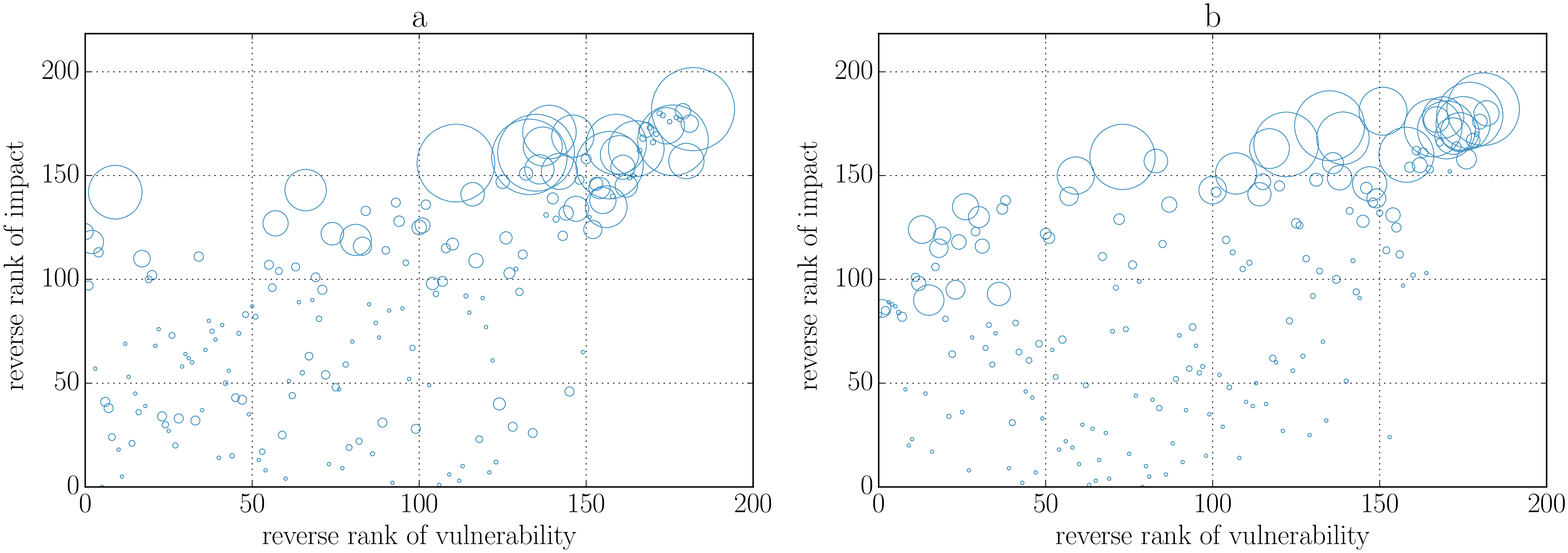}
\caption{\textbf{Scatter plot of impact and vulnerability (reverse) rankings in 2008 (panel a) and 2013 (panel b)}. An initial shock corresponding to a 0.5\% devaluation of its assets is applied to one bank at a time, and the experiment is repeated for each bank. The impact of a bank is measured as the relative equity loss experienced by the system when that bank is shocked. The vulnerability of a bank is its relative equity loss averaged over all the experiments. In addition, we average impact and vulnerability across a sample of 100 reconstructed networks. Finally, we build reverse ranking (i.e.\ in descending order) of both quantities, so that larger values on both axes correspond to more impactful and more vulnerable banks. Bubble size is proportional to the total assets of the corresponding bank. The most dangerous banks are also the most vulnerable.}
\label{fig:ranks}
\end{figure}

As a second scenario, we consider the case in which a single bank at a time is shocked, a shock still being a devaluation of its external assets by a relative amount $\alpha$, and the experiment is repeated for each bank. The idea is to decompose the systemic importance of a bank into its impact on the system and into its vulnerability with respect to shocks affecting other banks. We then proceed to define the impact of bank $i$ as the relative equity loss of the system when bank $i$ is shocked. Instead we take as a measure of its vulnerability the average of $h_i(t)$ over all the experiments. We then rank banks in descending order both in terms of impact and in terms of vulnerability and present the results for 2008 and 2013 and for $\alpha = 0.5\%$ in form of a scatter plot in Fig.\ \ref{fig:ranks}. We can see that the most dangerous banks, i.e.\ the banks having the largest impact on the system, are also the most vulnerable ones.

\section*{Discussion}
By iterating the balance sheet identity we derive an equation for the evolution of banks' equities. We then consider a general shock propagation mechanism in interbank networks so that the value of interbank assets of lenders depends on the level of distress of their creditors. The resulting dynamics is closely connected with the DebtRank algorithm recently introduced in the literature \cite{battiston2012debtrank} as an effective shock propagation dynamics and it provides a clear economic intuition for its dynamical variables, in terms of basic accounting principles. We prove that, in general, the original DebtRank gives a lower-bound for losses computed with our methodology, but, for a certain class of shocks, the two algorithms are equivalent on trees.
More importantly, we show that the capability of the system to amplify an initial shock depends only on the modulus of largest eigenvalue $\lambda_{\text{max}}$ of the matrix of interbank leverages: When $|\lambda_{\text{max}}|<1$, additional losses induced by subsequent rounds of the dynamics are attenuated over time. In contrast, when $|\lambda_{\text{max}}|>1$ a small shock will be amplified and cause at least one bank to default. 
This finding can be important from a regulatory perspective, as one could monitor the evolution of $\lambda_{\text{max}}$ over time to check if the system is entering the unstable regime.

To showcase our algorithm, we apply it to a system composed of 183 European publicly traded banks. We characterise the response of the network to different shock scenarios. Our analysis shows that the amplification of shocks due to interbank contagion consistently decreases from 2008 to 2013, and that in 2008 small shocks are enough for all banks to be significantly distressed. By performing stress tests in which banks are initially shocked one at a time, we are able to compute both the impact of a single bank on the system and its vulnerability to shocks initiated by other banks. From a systemic standpoint, it would be desirable that systemic impact and vulnerability were anti-correlated, so that the most dangerous bank are also the most robust, and vice versa. In fact, this does not happen: our analysis shows that the most dangerous banks are also the most vulnerable, meaning that systemic risk is concentrated in a few key players, which should  therefore be the objective of effective macroprudential regulation policies.

\section*{Methods}

\subsection*{Balance sheet basics}
A balance sheet summarises the financial position of a bank. It consists of assets, which have a positive economic value (e.g.\ stocks, bonds, cash), and liabilities, which are obligations to creditors (e.g.\ customers' deposits, other debits). The difference between the value of assets and liabilities is called equity, and the following (balance sheet) identity holds: $\text{assets} = \text{equity} + \text{liabilities}$. A bank is said to be solvent as long as its equity is positive. Once a bank is insolvent, even if it sold the entirety of its assets, it would not be able to repay its debts. As a consequence, we use insolvency as a proxy for default.

\subsection*{Model dynamics} 
The equation for the evolution of the cumulative relative loss of equity $h_i(t) = \left( E_i(0) - E_i(t) \right) / E_i(0)$ can be derived from the balance sheet identity. From \eqref{eq:bs_identity}, supposing that 
(i) external assets and liabilities do not change, (ii) interbank liabilities are at face value and also do not change, and (iii) interbank assets are marked-to-market, 
\begin{equation} \label{eq:delta_equity}
\begin{split}
E_i(t+1) - E_i(t) &= \sum_{j \in \mathcal{A}(t)} A_{ij}(t+1) - \sum_{j \in \mathcal{A}(t-1)} A_{ij}(t) \\
&= \sum_{j \in \mathcal{A}(t-1)} \left[A_{ij}(t+1) - A_{ij}(t) \right]  - \sum_{j \in \mathcal{A}(t-1) \setminus \mathcal{A}(t)} A_{ij}(t+1) \, ,
\end{split}
\end{equation}
where in the second line we have isolated a potential contribution coming from the nodes  that were active at time $t-1$, but became inactive at time $t$. Using \eqref{eq:shocks}, we see that the last term in \eqref{eq:delta_equity} vanishes, so that we have:
\begin{equation}
\begin{split}
E_i(t+1) - E_i(t) &= \sum_{j \in \mathcal{A}(t-1)} \frac{A_{ij}(t)}{E_j(t-1)} \left[ E_j(t) - E_j(t-1) \right] \\
&= \sum_{j \in \mathcal{A}(t-1)} \frac{A_{ij}(0)}{E_j(0)} \left[ E_j(t) - E_j(t-1) \right] \, ,
\end{split}
\end{equation}
where in the second line we have recursively applied \eqref{eq:shocks} and used $A_{ij}(1) = A_{ij}(0)$ (only equities change at time $t=1$, assets start to change at time $t=2$). We can now define the matrix $\tilde{\Lambda}$:
\begin{equation} \label{eq:l_tilde}
\tilde{\Lambda}_{ij}(t) = 
\begin{cases} 
\frac{A_{ij}(0)}{E_j(0)} &\mbox{if } j \in \mathcal{A}(t-1) \\
0 &\mbox{if } j \notin \mathcal{A}(t-1) \, 
\end{cases}
\end{equation}
and write the equation for the evolution of equity:
\begin{equation} \label{eq:evo_equity}
E_i(t+1) = \max \left[ 0, \, E_i(t) + \sum_{j=1}^N \tilde{\Lambda}_{ij}(t) \left[ E_j(t) - E_j(t-1) \right] \right] \, ,
\end{equation}
where the $\max$ accounts for the fact that once a bank defaults its equity cannot become negative. From \eqref{eq:evo_equity}, it easily follows that:
\begin{equation}
h_i(t+1) = \min \left[ 1, \, h_i(t) + \sum_{j=1}^N \Lambda_{ij}(t) \left[ h_j(t) - h_j(t-1) \right] \right] \, ,
\end{equation}
where $\Lambda_{ij}(t) = \tilde{\Lambda}_{ij}(t) E_j(0) / E_i(0)$, so that $\Lambda(t)$ can be interpreted as a \emph{reduced} interbank leverage matrix, where columns corresponding to banks defaulted up to time $t-1$ have been set equal to zero. As the equity of defaulted banks does not change anymore after reaching zero, the rows of the leverage matrix corresponding to defaulted banks can be set equal to zero too.

\subsection*{Relation to DebtRank} 
The original DebtRank \cite{battiston2012debtrank} has the following dynamics:
\begin{equation} \label{eq:dr}
\begin{split}
h_i(t+1) &= \min \left[ 1, \, h_i(t) + \sum_{\mathcal{A}^\prime(t)} W_{ij} h_j(t) \right] \\
&= \min \left[ 1, \, h_i(t) + \sum_{\mathcal{A}^\prime(t)} W_{ij} \left[ h_j(t) - h_j(t-1) \right] \right] \, ,
\end{split}
\end{equation}
where $W_{ij} = \min(1, \Lambda_{ij})$, and $\mathcal{A}^\prime(t) = \{j : h_j(t) > 0 \;\; \text{and} \;\; h_j(t-1) = 0\}$, and the last term in the second line can be added because it is always equal to zero. Let us note that the definition of $\mathcal{A}^\prime(t)$ implies a different stopping criterion. In fact, in the original DebtRank nodes propagate shocks only once, immediately after the shock has been received. In our setting, instead, they could propagate shocks until they default. There are two main differences with respect to \eqref{eq:evo_h}: (i) the summation in \eqref{eq:dr} involves less terms than the summation in \eqref{eq:evo_h} since $\mathcal{A}^\prime(t) \subseteq \mathcal{A}(t) \subseteq \mathcal{A}(t-1)$ (the set of active nodes becomes smaller and smaller as banks default); (ii) $W_{ij} < \Lambda_{ij}$, for all $i$ and $j$. As a consequence, \eqref{eq:dr} provides a lower bound to relative cumulative losses of equity computed with \eqref{eq:evo_h}. 

In order to understand the role of the network topology, let us focus our attention on a node $r$. From \eqref{eq:shocks} we see that a shock can reach $r$ only through the neighbours $r$ borrows from, which in turn can be reached by a shock only through the neighbours they borrow from. In other words, if a single node $s$ is shocked at some time $t$, the only possible way for $r$ to experience the effects of such shock (at later times) is that a path from $r$ to $s$ exists. Let us for a moment suppose that such path $r \rightarrow i_1 \rightarrow i_2 \ldots \rightarrow i_{p-1} \rightarrow s$ is unique (and of length $p$); then there will be also a unique path leading from any node $i_k$ to $s$. The shock will propagate to node $i_{p-1}$ at the time $t+1$, but, if no additional node is shocked, and since no additional paths exist between $s$ and $i_{p-1}$, the status of node $i_{p-1}$ will not change from time $t+1$ to time $t+2$. Similarly the status of node $i_{p-2}$ will change only at time $t+2$, and so on, until the shock reaches node $r$ at time $t+p$. The status of any node $i_k$ on the path will change only at one time step. As a consequence, the result will be the same as if each node were active only when reached for the first time by the shock, as in the original DebtRank. However, this is true only for the nodes $r$ such that a unique path connecting them to the only shocked node $s$ exists. If there are additional paths between $r$ and $s$ the shock will propagate also along those paths, resulting in additional losses at the node $r$. In particular this is trivially true if the subgraph of nodes reachable (backwards) from the shocked node $s$ is a tree. If more than a single node is shocked, and if $r$ is reachable (backwards) from more than one of them, then, even if the graph is a tree, the (cumulative) loss experienced by $r$ at the end \emph{could} be larger than if a stopping criterion \`{a} la DebtRank were used. In particular the loss will be larger if the paths are overlapping, while it will be equal if the paths are not overlapping.

\subsection*{Stability properties}
Let us assume for simplicity that no banks default during the whole evolution, so that $\Lambda$ is constant over time (see \eqref{eq:l_tilde}). Defining $\Delta h(t) = h(t) - h(t-1)$, \eqref{eq:evo_h} can be written in matrix notation:
\begin{equation} \label{eq:evo_delta_h}
\begin{split}
\Delta h(t+1) &= \Lambda \, \Delta h(t) \\
&= \Lambda^t \Delta h(1) = \Lambda^t h(1) \, , 
\end{split}
\end{equation}
as $\Delta h(1) = h(1) - h(0)$, and $h(0) = 0$. By summing over all the time steps up to $t+1$ one gets:
\begin{equation} \label{eq:evo_h_2}
h(t+1) = \sum_{s=0}^{t+1} \Delta h(s) = \sum_{s=0}^{t+1} \Lambda^s h(1) \, .
\end{equation}
$\Delta h = 0$ is always a fixed point of the map \eqref{eq:evo_delta_h}, and it is stable as long as the modulus of the largest eigenvalue $\lambda_{\text{max}}$ of $\Lambda$ is smaller than one, meaning that the dynamics will damp subsequent propagations of an initial shock over time. In this case the sum in \eqref{eq:evo_h_2} will asymptotically converge to:
\begin{equation}
h^\infty = \left( 1 - \Lambda \right)^{-1} h(1) \, .
\end{equation}
In contrast, if $|\lambda_{\text{max}}| > 1$, $\Delta h(t)$ will become increasingly larger, leading to the default of at least one bank, independently from the initial shock.

Eq.\ \eqref{eq:evo_delta_h} clearly describes the first stages of the dynamics, up to the first default. Nevertheless, since the reduced leverage matrix does not change between two subsequent defaults, \eqref{eq:evo_delta_h} also holds between one default and the next one, provided that $\Lambda$ is replaced with the correct reduced leverage matrix $\Lambda(t)$. As a consequence, the dynamics will remain explosive as long as the modulus of $\lambda_{\text{max}}(t)$, the largest eigenvalue of $\Lambda(t)$ is larger than one. As more and more banks default $|\lambda_{\text{max}}(t)|$ will eventually become smaller than one, and the dynamics will finally converge.

It should be noted that modifying the original DebtRank dynamics \eqref{eq:dr} by allowing banks to propagate shocks as long as their equity is positive would lead to a double-counting of losses. Let us suppose again for simplicity that no banks default and that $W = \Lambda$. Iterating \eqref{eq:dr} leads to $h(t+1) = \left( \mathbb{I} + \Lambda \right)^t h(1)$,  and this quantity is always larger than the one obtained from equation  \eqref{eq:evo_h_2}, i.e. $\sum_{s=0}^t \Lambda^s h(1)$.

\subsection*{Data}
For our analysis we use the same dataset used in \cite{Battiston2015}. Information on banks' balance sheets are taken from the Bureau Van Dijk Bankscope database for 183 European banks that were publicly traded between 2008 and 2013. From this data source we extract information about: equity, total assets, total liabilities, total interbank assets $\tilde A_{i}$ and total interbank liabilities $\tilde L_{i}$. For details about the handling of missing data, the reader should refer to the aforementioned reference \cite{Battiston2015}.

As mentioned in the main text, the procedure to reconstruct a matrix of interbank assets $A_{ij}$ develops in two steps: in the first we generate a binary adjacency matrix, which encodes the topology of the network. This is done via a fitness model \cite{musmeci2013bootstrapping}, conveniently modified for directed networks. A link from bank $i$ to bank $j$ is inserted with probability $p_{ij} = \frac{z x_i^{\text{out}}{x_j^{\text{in}}}}{1 + z x_i^{\text{out}}{x_j^{\text{in}}}}$, where the fitness values of each banks are computed as $x_i^{out} = \tilde A_{i}/ \sum_j\tilde A_{j}$ and $x_i^{in} = \tilde L_{i}/ \sum_j\tilde L_{j}$, and the parameter $z$ is fixed to attain the desired network density (the number of links in the network divided by the number of possible links). In this paper we have set $z$ so that the density of the network is $5\%$. We then draw 100 networks according to the probabilities $p_{ij}$. 
For each network thus obtained, we then proceed to assign weights $A_{ij}$ to the links. To this end we use the RAS algorithm \cite{upper2004estimating}. This consists in the iteration of a map whose $n$-th step is:
\begin{equation*}
\begin{split}
A_{ij}^{(n)} &= \frac{A_{ij}^{(n-1)}}{\sum_j A_{ij}^{(n-1)}} \tilde A_i \\
A_{ij}^{(n+1)} &= \frac{A_{ij}^{(n)}}{\sum_i A_{ij}^{(n)}} \tilde L_i \, .
\end{split}
\end{equation*} 
At convergence, the above iteration ensures that $\sum_j A_{ij} = \tilde A_i$ and $\sum_i A_{ij} = \tilde L_j$ for all banks. Obviously, one must have that $\sum_i \tilde A_i = \sum_i \tilde L_i$. Since this is not the case for our data, we rescale liabilities $\tilde L_i$ so that the above relation holds.

\bibliography{NewDebtRank}

\section*{Acknowledgments}
MB, SB, and GC acknowledge support from: FET Project SIMPOL nr.\ 610704, FET project DOLFINS nr.\ 640772, and FET IP Project MULTIPLEX nr.\ 317532. FC acknowledges support of the Economic and Social Research Council (ESRC)
in funding the Systemic Risk Centre (ES/K002309/1). SB acknowledges the Swiss National Fund Professorship grant nr.\ PP00P1-144689. We thank Stefano Gurciullo and Marco D'Errico for sharing their expertise on the dataset.

\section*{Author contributions statement}
MB, SB, FC, and GC discussed the original idea. MB and FC developed the conceptual framework. MB wrote the code and performed the numerics. MB, SB, FC, and GC contributed to the interpretation of results and to writing the manuscript.

\section*{Additional information}
\textbf{Competing financial interests.} The authors declare no competing financial interests. \textbf{Source code.} All the analysis has been developed in Python and is available upon request to the authors.

\end{document}